\documentclass[letter]{pasj02}
\Received{$\langle$reception date$\rangle$}
\Accepted{$\langle$acception date$\rangle$}
\Published{$\langle$publication date$\rangle$}
\usepackage{xcolor}
\usepackage[normalem]{ulem}

\usepackage{tabularx}
\begin{document}

\title{ An ALMA view of the Jet-Arc CO clouds toward the TeV $\gamma$-ray source HESS~J1023-575 and Westerlund~2; Evidence for the footprints of microquasar jets, the very powerful cosmic-ray accelerator in the Galactic disk}
\author{
Yasuo \textsc{Fukui},\altaffilmark{1,3}
Kisetsu \textsc{Tsuge},\altaffilmark{2,3,4,9}\altemailmark
Hidetoshi \textsc{Sano},\altaffilmark{3}
G.~R. \textsc{Bhuvana},\altaffilmark{2}
Rin I. \textsc{Yamada},\altaffilmark{3,4}
Ryoji \textsc{Matsumoto},\altaffilmark{5}
Yuta \textsc{Asahina},\altaffilmark{6}
Tsuyoshi \textsc{Inoue},\altaffilmark{7}
Tim Lukas \textsc{Holch},\altaffilmark{8}
Emma de~O\~{n}a \textsc{Wilhelmi},\altaffilmark{8}
}

\altaffiltext{1}{Department of Physics, Nagoya University, Furo-cho, Chikusa-ku, Nagoya 464-8601, Japan}
\altaffiltext{2}{Institute for Advanced Research, Gifu University, 1-1 Yanagido, Gifu 501-1193, Japan}
\altaffiltext{3}{Faculty of Engineering, Gifu University, 1-1 Yanagido, Gifu 501-1193, Japan}
\altaffiltext{4}{National Astronomical Observatory of Japan (NAOJ), National Institutes of Natural Sciences (NINS)2-21-1, Osawa, Mitaka, Tokyo 181-8588, Japan}
\altaffiltext{5}{Department of Physics, Faculty of Science, Chiba University, Inage-ku, Chiba 263-8522, Japan}
\altaffiltext{6}{University of Tsukuba, 1-1-1 Tennodai, Tsukuba, Ibaraki 305-8577, Japan}
\altaffiltext{7}{Department of Physics, Konan University, Okamoto 8-9-1, Higashinada-ku, Kobe 658-8501, Japan}
\altaffiltext{8}{Deutsches Elektronen-Synchrotron DESY, Platanenallee 6, 15738 Zeuthen, Germany}
\altaffiltext{9}{Institute for Advanced Research, Nagoya University, Furo-cho, Chikusa-ku, Nagoya 464-8601, Japan}

\email{tsuge.kisetsu.i2@f.gifu-u.ac.jp}

\maketitle

\begin{abstract}
The TeV $\gamma$-ray source HESS~J1023-575 (HESS~J1023 hereafter) is one of the {brightest} {H.E.S.S.} sources near the young massive cluster Westerlund~2. HESS~J1023 
shows a remarkable {positional} alignment with {the} Jet and Arc CO clouds on its eastern and western sides over 170 pc length. We have carried out sub-pc scale observations of the CO clouds with ALMA and have discovered that the clouds consist of numerous thin filamentary features of $\sim$0.5 pc width and 10--20 pc length {at distance of 7.5 kpc}, which are well aligned with the Jet-Arc axis. Based on the magneto-hydrodynamical model of microquasar jets launched from {the center of the $\gamma$-ray source} HESS~J1023{-575}, we present an interpretation that the thin filamentary clouds are the footprints of the microquasar jets on the H{\sc i} gas. The model also explains the dissimilar Jet vs. Arc clouds in terms of H{\sc i} density difference on each side. By using the density of the CO and H{\sc i} gas and the $\gamma$-ray luminosity, we have calculated the cosmic ray proton energy $W_{\rm p}$ to be 7$\times$10$^{48}$ erg under the hadronic scheme, which is ten times larger than those derived in the TeV $\gamma$-ray SNRs RX~J1713{.7-3946} and RX~J0852{.0-4622}. {It is likely that HESS~J1023 has been active over 1--10 Myr, which is significantly longer than the duration of cosmic ray acceleration of the SNRs.} HESS~J1023 is therefore an outstanding source of cosmic rays equivalent to at least 1000 SNRs, and is possibly the most powerful CR accelerator in the Galactic disk. {A high energy compact source in HESS~J1023, which is likely a Myr-old black hole or neutron star, remains veiled due to heavy extinction.}
\end{abstract}
\begin{figure*}[]
 \includegraphics[width=\linewidth]{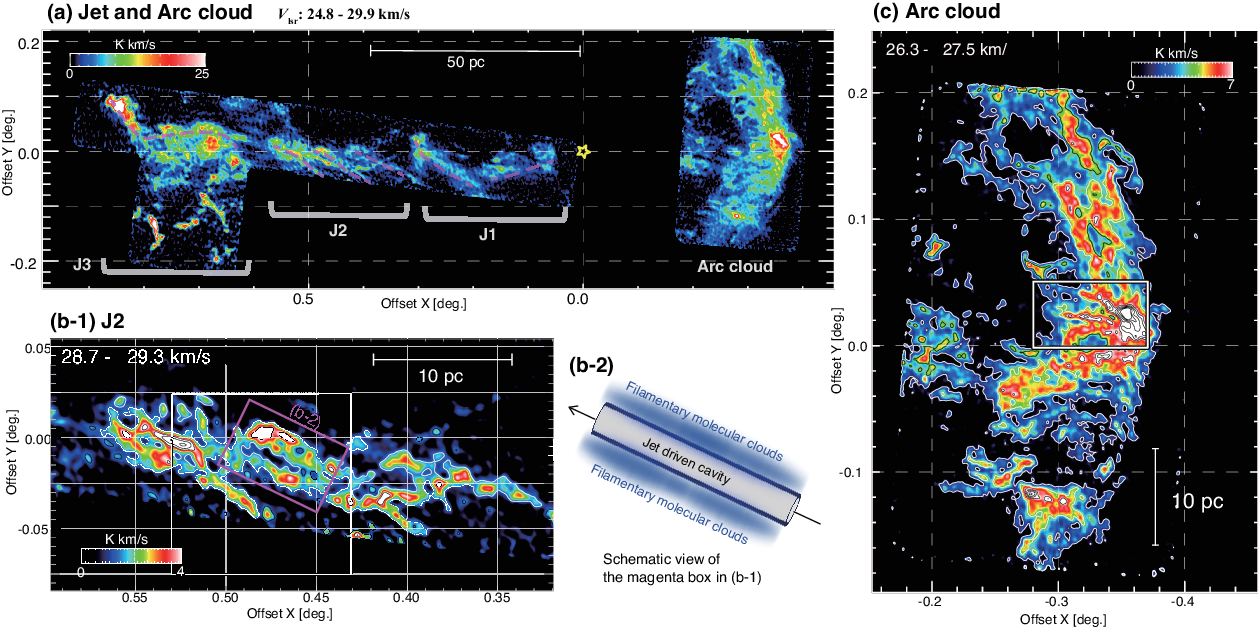}
 \caption{{Integrated intensity maps of the ALMA ACA+TP $^{12}$CO($J$=1--0) data.}The Galactic coordinates are rotated to align with the jet axis, $b(\mathrm{deg})=(-0.82\pm0.01)\,l(\mathrm{deg})+(234.0\pm3.3)$, derived from an intensity-weighted least-squares fit to the Jet and Arc clouds \citep{2014ApJ...781...70F}. The Offset X- and Y-axes are parallel and perpendicular to the jet axis, respectively, with HESS~J1023--575 (star) as the origin. The Jet cloud is divided into J1--J3 following \citet{2014ApJ...781...70F}. (a) Jet--Arc clouds integrated over $V_{\rm LSR}=24.8$--$29.9$ km s$^{-1}$. {Magenta guide lines mark representative filamentary ridges discussed in the text.} {(b-1) J2 integrated over $V_{\rm LSR}=28.7$--$29.3$ km s$^{-1}$. The magenta box indicates the region illustrated schematically in panel (b-2). (b-2) Schematic view of the jet-driven cavity and dense filamentary structures along its edges in the magenta box of panel (b-1).} (c) Arc cloud integrated over $V_{\rm LSR}=26.3$--$27.5$ km s$^{-1}$. {White boxes in panels (b-1) and (c) indicate the regions analyzed in Figure~2.}{Alt text: ALMA CO maps show filamentary molecular gas around a jet-like cavity and in the Arc cloud.}
}
 \label{fig:fig1}
\end{figure*}
\section{Introduction}
\vspace{-0.1cm}
Cosmic rays (CRs) are the most energetic particles in the Universe and their origin is a fundamental issue in astrophysics. It is thought that CRs with energy less than 10$^{15.5}$ eV are accelerated within the Galaxy, and SNRs are considered as the most promising CR accelerator. \citet{2021ApJ...915...84F} {(hereafter F21)} and \citet{2024ApJ...961..162F} {(hereafter F24)} derived that 50-60 \% of the total $\gamma$ rays consist of the hadronic component in two young TeV  $\gamma-$ray SNRs, RX~J1713.7-3946 (here after RX~J1713) and RX~J0852.0-4622 (here after RX~J0852). These authors utilized a methodology based on the interstellar proton density in order to quantify the hadronic component, and proved that the SNRs are a major accelerator of the Galactic CRs. In order to expand understanding the CR origin, it is essential to explore any other powerful mechanisms accelerating CRs. The young massive star clusters (Portegies Zwart et al. 2010) which have energetic winds are a possible CR accelerator, whereas recent studies showed that the contribution of the clusters in CR production is 1-10 \% of that of SNRs (e.g., \cite{2024NatAs...8..530P}). {On the other hand, recent theoretical studies suggested that SNR shocks embedded in compact massive clusters with strongly magnetized collective winds may accelerate CRs up to ultra-high energies, although the quantitative contribution of such environments to Galactic CRs is still under discussion (e.g., Vieu \& Reville 2023). Recent CO and H{\sc i} studies toward Westerlund~1 further suggested that the $\gamma$-ray environment around massive stellar clusters is likely complicated by mixing of stellar winds and possible SNR-related components, not allowing straightforward comparison with simple SNR cases (Sano et al. 2026).} 

Microquasars represented by SS433, V4641 Sgr, and GRS 1915+105 etc. are another promising source of CRs, which shows strong $\gamma$-rays and high energy jets (e.g., \cite{2025NSRev..12af496L}). 
Most recently, microquasars were suggested as 100 TeV $\gamma$-ray sources. $\gamma$-ray emission above 100 TeV from the extended jet of V4641 Sgr was detected by HAWC (\cite{2024Natur.634..557A}) and H.E.S.S. (\cite{2026A&A...706A...8A}). In addition, LHAASO {searched for UHE $\gamma$-ray emission from 12 black-hole microquasars and detected significant emission from five systems: SS 433, V4641 Sgr, GRS 1915+105, MAXI J1820+070, and Cygnus X-1} (\cite{2025NSRev..12af496L}). These recent observations indicate that microquasars are efficient CR accelerators which may produce CRs beyond 1 PeV ($>$ 10$^{15}$ eV). {However, high-energy emission has been detected in only a subset of known microquasar systems, and how their relativistic jets interact with and transfer energy to the surrounding ISM remains poorly understood. Detailed observational studies of jet--ISM interaction remain scarce, and even for the prototypical system SS433, such interaction has only recently been demonstrated (Sakemi et al. 2026).} {This is partly because spatially resolved jets are detected in only a limited number of systems and because their emission often shows strong time variability.} It is thus becoming an urgent issue to explore the properties of microquasars, in particular, concerning the CR acceleration budget (e.g., \cite{2025arXiv251001369K}).

\begin{table}
\caption{{Progress in understanding the origin of the Jet--Arc clouds.}}
\vspace{-0.2cm}
\label{tab:progress}
\centering
\footnotesize
\setlength{\tabcolsep}{3pt}
\renewcommand{\arraystretch}{0.95}
\begin{tabular}{p{0.27\columnwidth}p{0.70\columnwidth}}
\hline
Study & Key finding \\
\hline
Fukui et al.\ 2009 &
Jet and Arc clouds discovered; explosive or microquasar origin proposed. \\

Furukawa et al.\ 2014 &
Microquasar origin favored, but without direct evidence for jet--ISM interaction. \\

Asahina et al.\ 2014 &
MHD simulations demonstrated molecular-cloud formation
through jet--H\,{\sc i} interaction and cooling instability. \\

Asahina et al.\ 2017 &
3D MHD simulations reproduced the Jet--Arc morphology
through different H\,{\sc i} filling factors. \\
\hline
\end{tabular}
\end{table}
\begin{figure*}[tb]
\includegraphics[width=\linewidth]{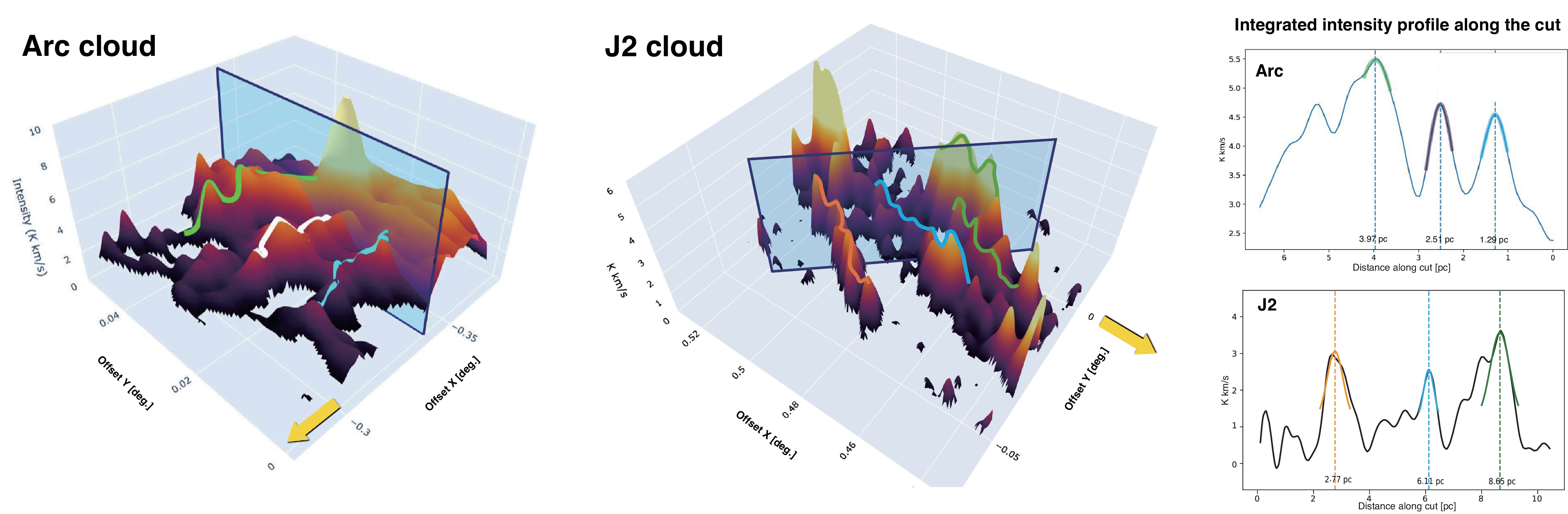}
\caption{Three-dimensional distributions and intensity profiles of the Arc and J2 clouds derived from $^{12}$CO($J$=1--0) data. The coordinate system is defined in the same manner as in Fig.~1, where the origin is set at HESS~J1023 and the axes are aligned with the jet direction. The left and middle panels show three-dimensional intensity distributions constructed using \texttt{plotly}. The vertical axis represents the integrated intensity. The left panel presents a zoomed view of the central region of the Arc cloud, while the middle panel shows the J2 cloud. Ridge structures are highlighted with color. The right panels show intensity profiles along the cuts indicated by the semi-transparent blue planes in the left panels. The horizontal axis represents the distance along the cut (pc), and the vertical axis shows the integrated intensity. Multiple peaks corresponding to the ridge structures are clearly identified. Each ridge is consistently color-coded between the three-dimensional maps and the profiles. Gaussian fitting is applied to each peak to measure the positions of the ridge structures. {Yellow arrows indicate the direction toward the central object HESS~J1023--575.} {Alt text:
Multiple parallel filaments are distributed with nearly regular spacing.}}
 \label{fig:fig2}
\end{figure*}

In the present paper we focus on a microquasar candidate HESS J1023-575 (HESS~J1023 hereafter, \cite{2007A&A...467.1075A,2011A&A...525A..46H}) and explore the possibility of a microquasar powering the gamma-ray source. HESS~J1023 was discovered in a field near a H{\sc ii} region RCW~49 ionized by a cluster Westerlund 2 which is located close to the center of gravity of the gamma-ray source HESS~J1023 with a 6$\arcmin$ separation. These authors suggested that Westerlund~2 is a possible accelerator of the CRs via stellar winds. The region is however toward the tangent of the Carina Arm and can be heavily contaminated by various unrelated clouds/objects (e.g., \cite{2007ApJ...665L.163D}). {The previous observational and theoretical studies relevant to the
origin of the Jet--Arc clouds are summarized in Table~1.}
\cite{2009PASJ...61L..23F} (Fukui09) discovered the Jet and Arc CO clouds with {the 4-m NANTEN2 radio telescope located in Atacama, Chile,} which show striking positional {alignment} with HESS~J1023.
\cite{2014ApJ...781...70F} (F14) made follow-up multi-line CO observations at 1.7--3.6 pc resolution with NANTEN2 and {the 22-m Mopra radio telescope in New South Wales, Australia} and argued that the clouds are associated with HESS~J1023 {region} at a kinematical distance of 7.5 kpc for the center of gravity velocity of the Jet-Arc CO clouds 26 km/s, which is separated by $\sim$20 km/s from the center of gravity velocity of the Westerlund 2 parent clouds (\cite{2009ApJ...696L.115F}; Furukawa09). Fukui09 and Furukawa09 proved that the Jet-Arc clouds are not physically linked with Westerlund 2 by showing no CO temperature rise toward Westerlund 2 (see Figure {12} of F14). F14 further showed localized high excitation CO spots, which are distributed far away up to $\sim$100 pc from Westerlund 2 (see Figure 1b); these spots show high temperatures of 20 K and/or high velocity dispersion exceeding 10 km s$^{-1}$, suggesting strong shock interactions in the Jet-Arc clouds. These observational signatures are located too far away to be excited by the cluster Westerlund 2. Fukui09 and Furukawa14 thus presented a scenario of the formation of the Jet--Arc clouds by high-energy events including a microquasar jet associated with HESS~J1023. However, no syncrotoron emission in radio or X-ray wavelength has not been detected 
{(e.g., Whiteoak \& Uchida 1997).}
The relativistic jets or compact objects were therefore not directly confirmed as in the case of SS433, and we call HESS~J1023 a microquasar candidate throughout this Letter (see Holch et al. 2021 and Holch et al. 2024 for an alternative scenario that considers the Jet and the Arc as separate objects, including HESS~J1024$-$583).



\vspace{-0.5cm}
\section{ALMA $^{12}$CO observations}

Most recently, the sensitive TeV $\gamma$-ray observations with {H.E.S.S.} successfully detected $\gamma$-rays coincident with the entire Jet-Arc clouds\footnote{
Prior to the discovery of the Jet-Arc clouds, the CO clouds at -10-20 km/s toward Westerlund 2 were observed with NANTEN2 at a resolution of 2.7 arcmin and were resolved into two CO clouds peaked at 4 km/s and 16 km s$^{-1}$, respectively (\cite{2009ApJ...696L.115F}). These authors associated the two clouds with Westerlund 2 based on their morphological correspondence at a distance of $5.4^{+1.1}_{-1.4}$ kpc. The association was confirmed via temperature rise of the clouds significantly heated by the cluster by \cite{2010ApJ...709..975O}. We note that the cluster Westerlund 2 has an age of 1.5 Myr, too young to form compact objects via a supernova, disfavoring a scenario that the Jet-Arc clouds were created by a cluster member.
} over 170 pc, and lent support for the high energy nature of the Jet and Arc clouds (Figures 5-26, 5-36 of \cite{Holch2021Exploring}). The $\gamma$-rays are likely of the hadronic origin because of the massive CO/H{\sc i} gas toward HESS~J1023 {with no synchrotron emission. X-ray pulsar wind nebulae have also been reported in the vicinity of HESS~J1023 \citep{2011A&A...525A..46H}.} Given an age 1-10 Myr of the Jet-Arc Clouds estimated by MHD model simulations (\cite{2017ApJ...836..213A}), an exciting possibility is emerging that HESS~J1023 may be a powerful CR accelerator in the Galaxy, even more efficient than an SNR.
To investigate the interaction between the relativistic jet and the ambient molecular gas in HESS J1023, we carried out ALMA $^{12}$CO($J$=1--0) observations toward the Jet and Arc clouds using the 7-m Atacama Compact Array (ACA) and the Total Power (TP) array. The observations were conducted in Cycle 12 under project code \#2025.1.00137.S (PI: K. Tsuge). {The raw data were calibrated using the standard ALMA calibration pipeline implemented in CASA (version 5.4.1-31). Imaging was performed with the CASA task \texttt{tclean}, and the ACA and TP images were combined using the \texttt{feather} task.} The synthesized beam size is $\sim14.4'' \times 11.7''$, corresponding to $\sim0.52 \times 0.43$ pc at a distance of 7.5 kpc. The velocity resolution is 0.3 km s$^{-1}$, and the typical rms noise level is 0.3 K per channel. The ACA+TP combination enables us to recover both compact and extended molecular structures without missing flux. The achieved spatial resolution is sufficient to resolve filamentary molecular structures at sub-pc scales and investigate possible signatures of relativistic jet interactions.



\vspace{-0.12cm}

\begin{figure*}[]
\includegraphics[width=\linewidth]{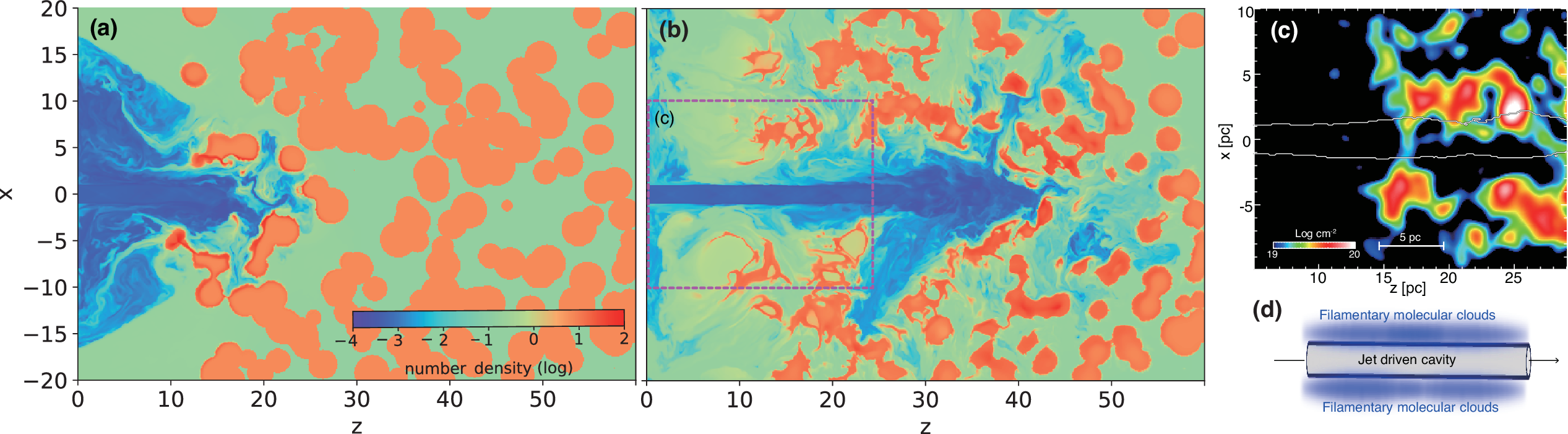}
\caption{{Density distributions from the three-dimensional magnetohydrodynamic simulation of the jet--ISM interaction based on the model of Asahina et al.\ (2017).}{Panels (a) and (b) show the number-density distributions at $t=1.67$ Myr and $6.5$ Myr after the onset of the jet injection, respectively.} The coordinate system is defined such that the jet propagates along the positive $z$-axis from the left side of the computational domain. {The color scale in panels (a) and (b) represents the logarithmic number density in units of cm$^{-3}$.} Filamentary structures produced by the interaction between the jet and the surrounding interstellar medium are visible in both panels. The low-density cavity seen near the center of {panel (b)} corresponds to the diameter of the propagating jet. {Panel (c) shows an enlarged view of the region enclosed by the magenta box in panel (b). The color scale represents the column density of cold gas with $T \leq 150$ K and line-of-sight velocities of $-1.0$ to $-0.4$ km s$^{-1}$. The original simulation data with a spatial resolution of 0.1 pc were Gaussian-smoothed to a spatial resolution of 0.5 pc, comparable to that of the ALMA observations. The white contours indicate gas with $v_z > 100$ km s$^{-1}$ and trace the propagating jet. The jet-driven low-density cavity and dense filamentary structures along its edges are clearly visible.} {Panel (d) schematically illustrates the structure identified in panel (c), in which dense filamentary gas structures are formed along the walls of the cavity excavated by the jet.} {In this model, the H{\sc i} clumps have a volume filling factor of 0.5.} The computational domain is $(40~{\rm pc}) \times (40~{\rm pc}) \times (60~{\rm pc})$ with a spatial resolution of 0.1 pc. {Alt text: Numerical simulations show a jet-driven low-density cavity surrounded by dense filamentary structures.}}
 \label{fig:fig3}
\end{figure*}
\vspace{-0.5cm}

\section{Jet and Arc CO clouds resolved with ALMA and a confrontation with the MHD model}
\subsection{ALMA Results}
The CO ($J$ = 1--0 and 2--1) distributions of the Jet-Arc clouds by F14 at 2-3 pc resolution show high temperatures up to 20 K in several locations. We here present the $^{12}$CO ($J$ = 2--1) distribution resolved at 0.5 pc with ALMA in Figure 1. Most remarkable are numerous thin filamentary features, which were resolved for the first time and were persistent over the Jet and Arc clouds. The typical width and length of these features are 0.5 pc and 20 pc, respectively, according to a detailed analysis with \texttt{FilFinder} by Tsuge et al. (2026 in preparation). The filamentary features run along the Jet cloud axis with regularity. 
{To quantify the spacing between the filamentary structures, we generated cuts perpendicular to the filament ridges across the J2 cloud, as illustrated in Figure~2. One representative cut is shown in Figure~2. Among the 10 approximately parallel cuts, five representative cuts located between Offset $\sim0.50$--$0.46$ deg along the Jet axis were adopted for the Gaussian fitting analysis.
The filament widths derived from the Gaussian fitting range from $0.27$ to $1.17$ pc, with an average value of $0.70 \pm 0.22$ pc.
The separations between adjacent ridges are derived to be $3.37 \pm 0.29$ pc between the outer and central ridges, and $2.11 \pm 0.35$ pc between the central and inner ridges.
Similar cuts were also generated for the Arc cloud. Gaussian fitting analyses were performed for three representative cuts. The derived ridge separations are $2.12 \pm 0.34$ pc between the outer and central ridges, and $1.81 \pm 0.25$ pc between the central and inner ridges. These results suggest that the filamentary structures in both the Jet and Arc clouds are distributed with quasi-regular spacings on scales of a few parsecs.} Such thin features have never been observed in molecular clouds as far as we are aware of, indicating that they are very rare.
\vspace{-0.7cm}
\subsection{The 3D MHD numerical simulations by Asahina et al. (2017) and their confrontation with the ALMA {results}}
\noindent{\bf MHD model}\\
\cite{2017ApJ...836..213A} made magneto-hydrodynamical (MHD) numerical simulations of the Jet-Arc clouds by incorporating a microquasar jet, and showed that the Jet-Arc clouds can be created by a SS433-type microquasar jets which interact with the ambient H{\sc i} gas. The usual inhomogeneous H{\sc i} gas consisting of the dense CNM (cold neutral medium) and the diffuse WNM (warm neutral medium) were assumed to have a mass ratio of $\sim$1:1. The CNM has a typical clump size of 0.1 pc and the WNM fills most of the volume between the CNM clumps. 
The findings of the simulations are two-fold. First, the microquasar jet drills the H{\sc i} into an evacuated tube of a few pc diameter associated with thin compressed CO walls which appear as thin filaments in projection. The ambient H{\sc i} is converted to H$_{2}$ via dust surface reactions under the high pressure of the jet in Myr. Second, in the 10--100 pc scale the twin microquasar jets create the Jet-like cloud in the east, and the Arc-like cloud in the west, where the difference in shape is due to difference of average H{\sc i} density by a factor of 2; i.e., in the eastern side low density of the CNM clumps allows the microquasar jet to propagate straight with small deceleration up to 60 pc, whereas in the western side the microquasar jet is strongly deflected and decelerated by the high density CNM clumps to form an arc-like shape with a radius of 50 pc. {To more clearly demonstrate the formation of the filamentary structures revealed by ALMA, we newly performed a simulation with the H{\sc i} clump filling factor increased from 0.2 to 0.5, while keeping all other model parameters of \citet{2017ApJ...836..213A}
unchanged. The increased filling factor enhances the jet--clump interaction, making the jet-driven displacement of the clumps and the resulting filamentary structures more clearly visible (Figure~3).} \\
\noindent{\bf Confrontation of the Model with the ALMA CO distribution}\\
{We compare the thin filamentary CO structures revealed by ALMA (Figures~1 and 2) with the MHD simulation in Figure~3. Figures~3a and 3b show the time evolution of the jet--ISM interaction, in which the propagating jet displaces and compresses the H{\sc i} clumps, forming a low-density cavity and dense filamentary structures. Figure~3c shows these structures at a spatial resolution comparable to the ALMA observations, revealing a jet-driven cavity bordered by dense filaments, similar to the morphology observed in J2 (Figure~1b-1).}
The jet can become multiple by deflection if it collides with the dense CNM clumps which are sufficiently massive (Asahina et al. 2017).

The large scale distribution of the CO clouds as denoted by J1,  J2,  J3, and Arc are reproduced in Figure 1a. We confirm that the mass of the Arc cloud within 50 pc from HESS~J1023 is 3$\times$10$^4$ M$_{\odot}$, and the mass of the eastern jet within 50 pc including J1 and J2 is 1.6$\times$10$^4$ M$_{\odot}$. So, the interacting CO mass ratio within the inner 50 pc is about 1:2 between the eastern and western sides, as is consistent with the Asahina et al.’s suggestion. In the western side, the jet is decelerated and terminated in the Arc cloud at 50 pc, while in the eastern side the jet propagates further up to 130 pc and interacts with more CNM clumps, being deflected spatially in a similar way to the Arc probably due to the large H{\sc i} mass located in the east at distance of 80-120 pc. This deflection causes broader CO linewidth of 10 km s$^{-1}$ in J3. We find that in the easternmost CO clump J3a even an infrared object is found (see Figure 17 of F14), which is likely a low mass young star formed in the shock compressed CO clump. The total length of the eastern Jet is 120 pc and that a radius of the arc is 40 pc. The total mass involved over a length of 170 pc is 8.4$\times$10$^4$ M$_{\odot}$.\\

\vspace{-0.5cm}
\noindent{\bf The timescale}\\
The time scale of the interaction can be estimated to be $\sim$10 Myr for the formation of a 50 pc jet like feature (Figure 10 of \cite{2017ApJ...836..213A}). This value depends on the assumed energy of the microquasar jets and the CNM density of the \cite{2017ApJ...836..213A} model that assumes that the microquasar jets have kinetic energy equivalent to 10 \% of that of SS433. If we assume the energy equal to the SS433 jet, ten times of what was assumed in \cite{2017ApJ...836..213A} the timescale above can become $\sim$1 Myr. So, we infer that the typical propagation timescale for the formation of the HESS~J1023 type jet is in a range of (1--10) Myr, depending on the jet energy and H{\sc i} density. More detailed jet interaction can be assessed from the ALMA image. One is the continuity of the thin filamentary features over the full length of the CO jet. The filamentary features show no apparent breaks, suggesting no strong break-off in the high energy jet. This suggests that the jet activity has been fairly steady although time resolution is coarse at 0.5 pc/50 pc $\times$ (1--10) Myr $\sim$ (10$^4$-10$^5$)yr, not sensitive to rapid time variation. 
At less than a sub-pc scale the interaction compresses the H{\sc i} to higher density like 1000 cm$^{-3}$, leading to formation of $H_2$ and CO via dust surface and gas phase reactions within Myr. It is likely that these shock interactions also heat up and excite turbulence which explains some warm spots of 20 K as well as of large velocity dispersions of a few to 10 km s$^{-1}$. So, the model provides a picture consistent with the observations. Because the jet is likely associated with magnetic field which causes the magnetic instability like the helical mode. {Aligned filamentary ridges, indicated by the magenta guide lines in Figure~1a, are seen at projected distances of 50--120 pc from HESS~J1023. Their gently curved morphology is consistent with molecular gas interacting with a helical jet subjected to MHD instability. A similar large-scale winding morphology was also identified in the lower-resolution CO observations of Fukui et al.\ (2009).}
Their gently curved morphology {is consistent with} CO gas interacting with a helical jet subjected to the MHD instability.
Another feature of concern is small systematic deviations of the angle of the CO jet-like features by $\sim$10 degrees from the direction of the $\gamma-$ray peak of HESS~J1023. We suggest that the proper motion of the H{\sc i} gas, generally a few km s$^{-1}$, over 100 pc can cause such displacements of a few pc in Myr after the interaction with the microquasar jet. We also note a trend that the $^{12}$CO intensity tends to become enhanced from J1 to J3. This may be explained by the elapsed time Myr after the shock heating, which is longer than the cooling time scale, 10$^4$--10$^5$ yr, since the interaction in J1 took place earliest in J1 and most recent in J3.
\vspace{-0.2cm}

{Combined with the MHD simulations, the quasi-regular filamentary structures and their characteristic spacing suggest that the ALMA CO observations may have resolved the dense cavity walls excavated by the microquasar jet.} 
{The multiple filamentary ridges with quasi-regular spacing newly resolved by ALMA in both the Jet and Arc clouds provide an additional observational constraint on their origin. This suggests that similar dynamical processes have acted on the molecular gas. Although the alternative interpretation in which the Jet and Arc clouds are unrelated structures (Holch et al. 2021, 2024) cannot be completely ruled out.}  
\vspace{-0.4cm}
\section{The $\gamma$-rays and Cosmic rays in HESS~J1023}
{{H.E.S.S.} collaboration promoted follow-up $\gamma$-ray observations of HESS~J1023–575 (\cite{2011A&A...525A..46H}), based on an expanded dataset with approximately twice the exposure compared to the earlier observations reported by \citet{2007A&A...467.1075A}.}
The most recent {H.E.S.S.} TeV $\gamma$-ray results were investigated (\cite{2024icrc.confE.778H,Holch2021Exploring}). The new {H.E.S.S.} $\gamma-$ray distribution shows unambiguously the component coincident with the CO Jet enveloped by H{\sc i}, which lends further support for that the Jet-Arc clouds are physically {connected} with {H.E.S.S $\gamma$-ray distribution}.

The $\gamma$-rays in HESS~J1023 prove that high energy particles of energy up to at least $\sim$10 TeV are accumulated within the $\gamma$-ray extent of HESS~J1023 {region}, where the $\gamma$-rays can be created by the hadronic or leptonic process. The origin of the $\gamma$ rays toward the CO Jet is yet to be pursued further. In HESS~J1023 synchrotron X-rays are not detected, indicating that high energy electrons may be not so significant as in case of {the synchrotron X-ray SNRs like RX~J1713 and RX~J0852.} The dense H{\sc i} gas surrounds the $\gamma$-ray distribution with a {24 pc} radius and the surrounding H{\sc i} gas favor the hadronic $\gamma$-ray production inside the CO Arc. {By adopting an average H{\sc i} density of 12 cm$^{-3}$ (F14), the cosmic ray proton energy $W_{\rm p}$ is expressed as follows {(Bosch-Ramon et al.~2005)},} 
{
\begin{equation}
\begin{split}
W_{\rm p} \ (10-100 \ \text{TeV}) \sim 7 \times 10^{48} \ \text{erg} \left( \frac{t_{\rm pp}}{5\times10^{6} \ \text{yr}} \right) \\
\times \left[ \frac{L_{\gamma} \ (1-10 \ \text{TeV})}{4.5\times10^{34} \ \text{erg s}^{-1}} \right].
\end{split}
\end{equation}
}

{Here, $t_{\text{pp}}$ is the cooling time-scale due to the pp-interaction and $L_{\gamma}$ is the $\gamma-$ray luminosity in the energy range of 1$-$10 TeV derived at a distance of 7.5 kpc.} We thus calculate $W_{\rm p}$ to be {7$\times$10$^{48}$ erg}. The high energy particles are lost by cooling and diffusion without energy injection. The cooling time scale and the diffusion time scale of the CR protons in HESS~J1023 are given by
{
\begin{equation}
t_{\rm pp} = 5\times10^6 \left( \frac{n}{12 \ \text{cm}^{-3}} \right)^{-1} \ {\rm yr}
\end{equation}}
and
{
\begin{equation}
\begin{split}
t_{\rm diffusion}= 2\times10^{4} \ \text{yr} \left( \frac{\chi}{0.01} \right)^{-1} \left( \frac{R}{24 \ \text{pc}} \right)^{2} \\
\times \left( \frac{E_{\rm p}}{10 \ \text{TeV}} \right)^{-0.5} \left( \frac{B}{10 \ \mu G} \right)^{0.5}
\end{split}
\end{equation}}
respectively {(Gabici et al. 2009)}.

{Here, $n$ is the density of the target protons within the radius of $\gamma-$ray extent $R$,  $\chi$ is the deviation from typical diffusion co-efficient in the Galaxy, $E_{\text{p}}$ is the CR proton energy and $B$ is the magnetic field strength. The reference values for these parameters are adopted from F14. We thus estimate the cooling and diffusion time scales to be 5$\times$10$^6$ yr and 2$\times$10$^4$ yr, respectively. Therefore, the CR energy is being lost mainly by diffusion over time scales of $\sim10^4$ yrs, needs an energy injection of $\sim$ 2$\times$10$^{37}$ erg s$^{-1}$ (7$\times$10$^{48}$ erg / 3$\times$10$^{11}$ s).} This energy can be supplied by the microquasar by mass accretion from a high mass OB star onto a compact star, either a black hole or neutron stars, inside HESS~J1023.

F21 and F24 calculated $W_{\rm p}$ (SNR) $\sim$8$\times$10$^{47}$ erg both in the two young SNRs RX~J1713 and RX~J0852 which are established as CR proton accelerators. {These two SNRs provide unique benchmark objects because the hadronic and leptonic $\gamma$-ray components have been quantitatively separated using observational constraints from the interstellar proton distribution and the X-ray synchrotron emission (F21, F24), allowing robust estimates of the CR proton energy $W_{\rm p}$. Similar estimates are not yet available for microquasars such as SS433 because the hadronic and leptonic $\gamma$-ray components have not been quantified separately.} $W_{\rm p}$ in HESS~J1023 is therefore an order of magnitude larger than that derived in the SNRs. The time scales of the active phase of CR acceleration in the SNRs are a few times 1000 years (see \cite{2015ARNPS..65..245F}). In case of HESS~J1023, the timescale of active microquasar jets is estimated to be 1--10 Myr from the MHD model in Section 3 (\cite{2017ApJ...836..213A}). The total supply of the CR protons is therefore given by a ratio 10 $\times$ (1--10 Myr) / (1000 yr) = 1000--10000 relative to that by a SNR
This indicates that HESS~J1023 {region} can be a cosmic accelerator which is equivalent to at least 1000 SNRs over its lifetime. The remaining factor concerned is a ratio of microquasars vs. SNRs, which is known as $\sim$20:300, possibly suggesting that microquasars can overwhelm SNRs in the whole production of Galactic cosmic rays.

\vspace{-0.6cm}
\section{Conclusions}
We carried out ALMA observations of a microquasar candidate HESS~J1023 in the $^{12}$CO ($J$ = 2--1) transition. 
The main conclusions are summarized as follows:

\begin{enumerate}
\item 
We have resolved the Jet-Arc CO clouds into numerous thin filamentary clouds of 0.5 pc width and 20--30 pc length. The filamentary clouds are remarkable features aligned along the Jet-Arc axis and are spaced by {1}--3 pc with each other. We present an interpretation that the filamentary clouds are the footprints of the microquasar jets assumed launched from the compact stellar object, either a black hole or a neutron star, which drilled the pre-existent H{\sc i} gas into evacuated tubes with dense gas walls.  The dense gas wall is thin and appears as the thin CO filamentary clouds. The microquasar is currently not directly observed but is strongly favored by the present discovery, {and is independently supported by the presence of the {H.E.S.S.} TeV $\gamma-$ray feature revealed by the recent detection of the jet-like extension of {HESS~J1023-575 or HESS~1024-583}}. 

\item 
Confrontation of the filamentary clouds with magneto-hydrodynamical numerical simulations shows the following aspects of the interactions: 
In a 10 pc scale, the twin microquasar jets create different shapes, i.e. straight jet and crescent arc, depending on the initial density of the H{\sc i} gas. 
{As is consistent with a mass ratio of 1:2 between the eastern and western sides within 60 pc.}
\item 
The timescale of the microquasar jet is suggested to be (1--10) Myr for microquasar jets having 0.1-1.0 dynamical luminosity of SS433 based on the MHD simulations. The observed uniformity of the thin filamentary CO clouds suggests that the physical properties of the microquasar jets do not vary significantly in time, indicating steadiness of the compact driving source. 
\item 
Cosmic ray proton energy $W{\rm p}$ in HESS~J1023 is an order of magnitude larger than that derived in the young TeV $\gamma-$ray SNRs. The time scales of the active phase of CR acceleration in the SNRs are a few times 1000 years. In case of HESS~J1023, the timescale of active microquasar jets is estimated to be 1-10 Myr, and the total supply of the CR protons is given by a ratio 1000-10000 relative to that of a SNR.  This indicates that HESS~J1023 {region} can be a cosmic accelerator which is equivalent to at least 1000 SNRs over its lifetime, suggesting an important role of microqusars in cosmic ray acceleration.

\end{enumerate}
\vspace{-0.6cm}
\begin{ack}
This paper makes use of the following ALMA data: ADS/JAO.ALMA \#2025.1.00137.S. ALMA is a partnership of ESO (representing its member states), NSF (USA) and NINS (Japan), together with NRC (Canada), NSTC and ASIAA (Taiwan), and KASI (Republic of Korea), in cooperation with the Republic of Chile. The Joint ALMA Observatory is operated by ESO, AUI/NRAO and NAOJ. This work was also supported by JSPS KAKENHI grant Nos. 25K17435 (K. Tsuge), 21H01136 (HS) and 24H00246 (HS). This work was supported by the Tokai Pathways to Global Excellence (T-GEx), part of the MEXT Strategic Professional Development Program for Young Researchers, and by the establishment of university fellowships toward the creation of science technology innovation (Grant Number: JPMJFS2138). {NAOJ ALMA Scientific Research Grant Code 2023-25A also supported this work.}
\end{ack}


\vspace{-0.4cm}
\begin{thebibliography}{}

\bibitem[Acharyya et al.(2026)]{2026A&A...706A...8A} Acharyya, A., Aharonian, F., Ashkar, H., et al.\ 2026, \aap, 706, A8. doi:10.1051/0004-6361/202557532
\bibitem[Aharonian et al.(2007)]{2007A&A...467.1075A} Aharonian, F., Akhperjanian, A.~G., Bazer-Bachi, A.~R., et al.\ 2007, \aap, 467, 3, 1075. doi:10.1051/0004-6361:20066950
\bibitem[Alfaro et al.(2024)]{2024Natur.634..557A} Alfaro, R., Alvarez, C., Arteaga-Vel{\'a}zquez, J.~C., et al.\ 2024, \nat, 634, 8034, 557. doi:10.1038/s41586-024-07995-9
\bibitem[Asahina et al.(2014)]{2014ApJ...789...79A} Asahina, Y., Ogawa, T., Kawashima, T., et al.\ 2014, \apj, 789, 1, 79. doi:10.1088/0004-637X/789/1/79
\bibitem[Asahina et al.(2017)]{2017ApJ...836..213A} Asahina, Y., Kawashima, T., Furukawa, N., et al.\ 2017, \apj, 836, 2, 213. doi:10.3847/1538-4357/aa5c86
\bibitem[Bosch-Ramon et al.(2005)]{2005A&A...432..609B} Bosch-Ramon, V., Aharonian, F.~A., \& Paredes, J.~M.\ 2005, \aap, 432, 2, 609. doi:10.1051/0004-6361:20041794


\bibitem[Dame(2007)]{2007ApJ...665L.163D} Dame, T.~M.\ 2007, \apjl, 665, 2, L163. doi:10.1086/521363

\bibitem[Fukui et al.(2009)]{2009PASJ...61L..23F} Fukui, Y., Furukawa, N., Dame, T.~M., et al.\ 2009, \pasj, 61, L23. doi:10.1093/pasj/61.4.L23

\bibitem[{Fukui et al.(2021)}]{2021ApJ...915...84F}
{Fukui}, Y., {Sano}, H., {Yamane}, Y., et~al. 2021, \apj, 915, 84,
  \dodoi{10.3847/1538-4357/abff4a}

\bibitem[{Fukui et al.(2024)}]{2024ApJ...961..162F}
{Fukui}, Y., {Aruga}, M., {Sano}, H., {Hayakawa}, T., {Inoue}, T.,
  {Rowell}, G., {Einecke}, S., \& {Tachihara}, K.
  2024, \apj, 961, 162,
  \doi{10.3847/1538-4357/ad0da3}


\bibitem[Funk(2015)]{2015ARNPS..65..245F} Funk, S.\ 2015, Annual Review of Nuclear and Particle Science, 65, 245. doi:10.1146/annurev-nucl-102014-022036  

\bibitem[Furukawa et al.(2009)]{2009ApJ...696L.115F} Furukawa, N., Dawson, J.~R., Ohama, A., et al.\ 2009, \apjl, 696, 2, L115. doi:10.1088/0004-637X/696/2/L115

\bibitem[Furukawa et al.(2014)]{2014ApJ...781...70F} Furukawa, N., Ohama, A., Fukuda, T., et al.\ 2014, \apj, 781, 2, 70. doi:10.1088/0004-637X/781/2/70

\bibitem[Gabici et al.(2007)]{2007Ap&SS.309..365G} Gabici, S., Aharonian, F.~A., \& Blasi, P.\ 2007, \apss, 309, 1-4, 365. doi:10.1007/s10509-007-9427-6
\bibitem[Gabici et al.(2009)]{2009MNRAS.396.1629G} Gabici, S., Aharonian, F.~A., \& Casanova, S.\ 2009, \mnras, 396, 3, 1629. doi:10.1111/j.1365-2966.2009.14832.x

  
\bibitem[H.~E.~S.~S. Collaboration et al.(2011)]{2011A&A...525A..46H} H.~E.~S.~S. Collaboration, Abramowski, A., Acero, F., et al.\ 2011, \aap, 525, A46. doi:10.1051/0004-6361/201015290
\bibitem[{Holch(2021)}]{Holch2021Exploring}
Holch, T.~L. 2021, PhD thesis, Humboldt-Universit\"{a}t zu Berlin, doi:10.18452/22390

\bibitem[Holch et al.(2024)]{2024icrc.confE.778H} Holch, T.~L., Hess, Aharonian, F., et al.\ 2024, 38th International Cosmic Ray Conference, 778. doi:10.22323/1.444.0778

  

\bibitem[Kaci et al.(2025)]{2025arXiv251001369K} Kaci, S., Giacinti, G., Aharonian, F., et al.\ 2025, arXiv:2510.01369. doi:10.48550/arXiv.2510.01369

  
\bibitem[Lhaaso Collaboration et al.(2025)]{2025NSRev..12af496L} Lhaaso Collaboration, Cao, Z., Aharonian, F., et al.\ 2025, National Science Review, 12, 12, nwaf496. doi:10.1093/nsr/nwaf496
\bibitem[Ohama et al.(2010)]{2010ApJ...709..975O} Ohama, A., Dawson, J.~R., Furukawa, N., et al.\ 2010, \apj, 709, 2, 975. doi:10.1088/0004-637X/709/2/975

\bibitem[Peron et al.(2024)]{2024NatAs...8..530P} Peron, G., Casanova, S., Gabici, S., et al.\ 2024, Nature Astronomy, 8, 530. doi:10.1038/s41550-023-02168-6
\bibitem[Portegies Zwart et al.(2010)]{2010ARA&A..48..431P} Portegies Zwart, S.~F., McMillan, S.~L.~W., \& Gieles, M.\ 2010, \araa, 48, 431. doi:10.1146/annurev-astro-081309-130834
\bibitem[Sakemi et al.(2026)]{2026ApJ..1004L..18S} Sakemi, H., Sano, H., Fukui, Y., et al.\ 2026, \apjl, 1004, 1, L18. doi:10.3847/2041-8213/ae736b
\bibitem[Sano et al.(2026)]{2026ApJ..1004...12S} Sano, H., Fukui, Y., Fujimori, S., et al.\ 2026, \apj, 1004, 1, 12. doi:10.3847/1538-4357/ae6337

\bibitem[Shapiro \& Teukolsky(1983)]{1983bhwd.book.....S} Shapiro, S.~L. \& Teukolsky, S.~A.\ 1983, . doi:10.1002/9783527617661
\bibitem[Vieu \& Reville(2023)]{2023MNRAS.519..136V} Vieu, T. \& Reville, B.\ 2023, \mnras, 519, 1, 136. doi:10.1093/mnras/stac3469
\bibitem[Whiteoak \& Uchida(1997)]{1997A&A...317..563W} Whiteoak, J.~B.~Z. \& Uchida, K.~I.\ 1997, \aap, 317, 563. 


\end{thebibliography}
\end{document}